\documentclass[twocolumn,superscriptaddress,showpacs,showkeys,preprintnumbers,amsmath,amssymb,prd]{revtex4-1}
\usepackage{graphicx}% Include figure files
\usepackage{dcolumn}% Align table columns on decimal point
\usepackage{bm}% bold math
\usepackage{amsmath,amssymb}

\begin{document}

\preprint{APS/123-QED}

\title{Addressing  the missing matter problem in galaxies through a new fundamental 
gravitational radius}

\author{S. Capozziello}
\email[Corresponding author:]{capozziello@na.infn.it}
\affiliation{Dipartimento di  Fisica "E. Pancini",  Universit\`{a} di Napoli "Federico II",
 Compl. Univ. di Monte S. Angelo, Edificio G, Via Cinthia, I-80126, Napoli, Italy,}
\affiliation{Istituto Nazionale di  Fisica Nucleare (INFN) Sez. di 
Napoli, Compl. Univ.di Monte S. Angelo, Edificio G, Via Cinthia, 
I-80126, Napoli, Italy,} \affiliation{Gran Sasso Science Institute,  Viale F. Crispi, 7, I-67100,
L'Aquila, Italy.}

\author{P. Jovanovi\'{c}}
\affiliation{Astronomical Observatory, Volgina 7, P.O. Box 74, 
11060 Belgrade, Serbia.}

\author{V. Borka Jovanovi\'{c}}
\affiliation{Atomic Physics Laboratory (040), Vin\v{c}a Institute of
Nuclear Sciences, University of Belgrade, P.O. Box 522, 11001
Belgrade, Serbia.}

\author{D. Borka}
\affiliation{Atomic Physics Laboratory (040), Vin\v{c}a Institute of 
Nuclear Sciences, University of Belgrade, P.O. Box 522, 11001 
Belgrade, Serbia.}

\date{\today}

\begin{abstract}
We demonstrate that the existence of a Noether symmetry in $f(R)$ theories of gravity  gives rise to 
a further  gravitational radius, besides the standard   Schwarzschild one,  determining the dynamics at galactic scales. By this 
feature, it is possible to explain the baryonic Tully--Fisher 
relation and the rotation curve of gas-rich galaxies without the 
dark matter hypothesis.
\end{abstract}

%\pacs{pacs1, pacs2, pacs3, pacs4}

% PACS, the Physics and Astronomy Classification Scheme.

\maketitle

\section{Introduction}
\label{sec_intro}

The standard $\Lambda\mathrm{CDM}$ cosmological model predicts that about 70\% 
of ''mass-energy budget'' of the Universe is composed of dark energy 
($\Omega_\Lambda\approx0.69$), about 30\% is the total mass of the Universe 
($\Omega_m\approx 0.31$) which is dominated with more than 5/6 by dark matter 
($\Omega_c\approx 0.26$), while ordinary baryonic matter constitutes only less 
than 1/6 ($\Omega_b\approx 0.05$) of the total mass (\citet{plnk16}). Flat 
rotation curves of spiral galaxies, together with discrepancy between the 
observed velocity dispersion and mass density of luminous matter in the 
clusters of galaxies, are usually assumed as the most significant observational 
evidences for the existence of dark mater (see e.g \citet{desw17} and 
references therein). In order to account for the flat rotation curves of spiral 
galaxies, dark matter should dominate the total mass of the galaxy and should 
be concentrated in the outer baryonic regions of galactic disks, as well as in 
the surrounding haloes. Thus, according to $\Lambda\mathrm{CDM}$, the baryonic 
matter in spiral galaxies is mixed with and embedded in non-baryonic dark 
matter, causing rotation velocities of the visible matter within the disk that 
are constant or increasing with disk radius. However, very recently 
\citet{genz17} studied the rotation curves in a sample of massive 
star-forming galaxies at the peak of cosmic galaxy formation 10 billion 
years ago, and found that in the case of 6 such galaxies at redshifts $z\sim 
0.9-2.4$ the rotation velocities are not constant, but decrease with radius.
Besides, the dark matter fractions near the half-light radius for all galaxies 
were modest or negligible. This indicated that a large fraction of the massive 
high-redshift galaxies were strongly baryon-dominated, with much less dark 
matter than in the local Universe, as well as that such baryon domination 
increased with redshift (\citet{genz17}). These results are in discrepancy with 
$\Lambda\mathrm{CDM}$ predictions obtained by cosmological galaxy formation 
simulations, according to which dark matter should play a key role in formation 
of the galaxies and their clusters because, in a $\Lambda\mathrm{CDM}$ 
Universe, dark matter first clumps and forms haloes and then galaxies form at 
the centres of these dark matter haloes (see e.g. \citet{spr06}).

On the other hand, weak lensing observations of merging galaxy cluster 1E 
0657-558 (the Bullet Cluster), were claimed to represent the first direct 
detection of dark matter (\citet{clow06}). Such claim was based on the inferred 
spatial offset of the center of the total mass of this cluster from the center 
of its baryonic mass peaks, which allegedly could be explained only by cold 
dark matter hypothesis. However soon afterwards, \citet{mahd07} performed a 
weak lensing study of the cluster merger known as Abell 520 (the Train Wreck 
Cluster) and found a massive dark core, i.e. a mass peak without galaxies which 
coincided with  the  gas. This could not be satisfactorily explained within the 
current collisionless dark matter paradigm since dark matter, being 
non-interacting, is expected to always produce a peak offset in the maps 
obtained by weak lensing mass reconstruction. Therefore, in a sense, this was a 
counterexample to the Bullet Cluster and represented a challenge to cold dark 
matter hypothesis which is one of the cornerstones of $\Lambda\mathrm{CDM}$. 
There was an attempt to dispute the existence of this dark core 
(\citet{clow12}), but more recent detailed observational studies of Abell 520 
by \citet{jee12,jee14} confirmed its existence. Moreover, \citet{lee10} 
demonstrated that the Bullet Cluster, initially considered as an evidence, 
actually represents a problem for $\Lambda\mathrm{CDM}$. Namely, high initial 
infall velocities of subclusters just before their collision, which are required 
to explain the observed physical and morphological properties of the Bullet 
Cluster, are incompatible with the much lower infall velocities predicted by 
$\Lambda\mathrm{CDM}$ (see \citet{lee10} and references therein).

As a consequence, these results indicated that either some further constraints 
on dark matter interaction should be set, such as the non-gravitational forces 
acting on dark matter (see e.g. \citet{harv15}), or an alternative explanation 
by modified gravity and without dark matter should be preferred.

Regarding the latter approach it was shown that MOND is able to successfully 
explain the dynamics of galaxies outside of galaxy clusters and a recently 
discovered tight relation between the radial acceleration inferred from their 
observed rotation curves and the acceleration due to the baryonic components of 
their disks (\citet{mcg16}). In the case of the Bullet Cluster and other 
galaxy clusters, MOND can reduce the amount of unseen matter to a factor of 
two, leaving the possibility that it could be non-luminous ordinary baryonic 
matter, rather than cold dark matter (see e.g. \citet{fama12}). It was also 
shown that TeVeS, the relativistic version of MOND, is able to resemble the weak 
lensing density maps of the Bullet Cluster by a multicentred baryonic system 
(\citet{ang06}). Some other theories of modified gravity which give the weak 
field point particle gravitational potential with Yukawa correction term are 
also able to explain both flat rotation curves and the Bullet Cluster (see e.g. 
\citet{brow07}). Recently, \citet{isr16} showed that generalized 
Scalar-Tensor-Vector-Gravity (STVG) theory can also explain dynamics of both 
merging galaxy clusters, Abell 520 and the Bullet Cluster, without dark matter.
This approach, however, requires modification of General Relativity and 
derivation of alternative field equations, as well as their cosmological 
reformulation.

In addition to the above  arguments,  it is well known that $f(R)$ theories of gravity gives rise, besides 
the standard Schwarzschild radius, to an additional ''characteristic 
length'' \cite{salv11,annalen,odintsov}. Such a feature has a 
fundamental nature and is related to the further degrees of freedom 
emerging as soon as $f(R)\neq R$, where $R$ is the Ricci curvature 
scalar \cite{starobinsky}. From the  conformal transformation point of view, being 
$f(R)$ gravity equivalent to a scalar-tensor theory, this new 
gravitational length is related to the effective mass emerging from 
the potential describing the dynamics of the scalar field. The 
paradigm is that being General Relativity (GR) a second-order theory in 
metric description, one has only a characteristic gravitational 
radius (the Schwarzschild radius). $f(R)$ gravity, being a 
fourth-order theory, has two characteristic gravitational radii and, 
in general, any theory of gravity of order $2N$ has $N$ 
characteristic radii that emerge in the weak field limit 
\cite{schmidt}.

These features can be of extreme interest in order to fix the size 
and the morphology of self-gravitating systems: in particular, the  
rotation curves of galaxies \cite{capo07,frig07, salucci}. The 
further gravitational radius emerges in any $f(R)$ model and, in 
particular, is related to the conserved quantities derived from 
the Noether symmetries in any power-law function of $R$. As shown in  
\cite{arturo, bern11}, for $f(R)\propto R^n$, where $n \in 
\mathbb{R}$, being $\mathbb{R}$ the set of real numbers, a Noether 
symmetry $\Sigma_0$ exists for any $n$. Specifically, 
$\displaystyle{\Sigma_0\equiv\frac{2GM}{c^2}}$, i.e. the 
Schwarzschild radius, is obtained for $n=1$, as it has to be in 
GR. For $n\neq1$, a further characteristic length emerges besides 
the Schwarzschild radius (see \cite{arturo} for 
details).  

Here, we want to show that this new characteristic radius can 
account for the baryonic Tully--Fisher (BTF) relation of gas-rich 
galaxies. In this perspective, the dark matter hypothesis is not 
necessary for the stability and the dynamics of such systems. In 
other words, galactic self-gravitating systems could be explained, 
in principle, only by gravitational interaction and baryons.

The paper is organized as follows: in Sec.~\ref{sec_theory} we give 
the basic ingredients of $f(R)$ gravity; in the Sec.~\ref{sec_btf},  
the observed BTF relation of gas-rich galaxies is derived in the 
framework of $f(R)$ gravity; we discussed the new gravitational 
radius depending on the mass of the system. Such an effective radius 
is straightforwardly related to the Modified Newtonian Dynamics 
(MOND) acceleration constant $a_0$ \cite{milg83}. Discussion and 
conclusions are drawn in Sec.~\ref{sec_concl}. In Appendix 
A, we sketch  the Noether Symmetry Approach by which it is possible 
to derive the further gravitational radius.

\section{Theory}
\label{sec_theory}

Let us start from the action
\begingroup
\setlength{\abovedisplayskip}{0pt}
\setlength{\belowdisplayskip}{0pt}
\begin{equation}
A = \int {d^4 x \sqrt{-g} \left[ f\left( R \right) + {\cal L}_m \right]},
\label{equ01}
\end{equation}
\endgroup

\noindent where $f(R)$ is a generic function of the Ricci curvature 
scalar and ${\cal L}_m$ is the standard matter Lagrangian. The field 
equations are
\begin{eqnarray}
R_{\mu \nu} - \frac{1}{2} g_{\mu \nu} R& = & \displaystyle{\frac{1}{f'(R)}}
\displaystyle{\Bigg \{ \frac{1}{2} g_{\mu \nu} \left [ f(R) - R
f'(R) \right ] + f'(R)_{; \mu \nu}} \nonumber \\ ~ & - &
\displaystyle{g_{\mu \nu} \Box{f'(R)} \Bigg \}} +
\displaystyle{\frac{T^{(m)}_{\mu \nu}}{f'(R)}} \label{eq:f-var2}
\end{eqnarray}
in a Einstein-like form. Primes denote derivative with respect to
$R$. The  terms ${f'(R)}_{; \mu \nu}$ and $\Box{f'(R)}$ give rise to 
fourth order derivatives in $g_{\mu \nu}$. Let us consider the  power\,-\,law case

\begin{equation}
f(R) = f_0 R^n \label{eq: frn}
\end{equation}
with $f_0$ a dimensional constant. Taking into account 
the gravitational field generated by a pointlike source and solving 
the field equations (\ref{eq:f-var2}) in the vacuum case, we 
write the  metric as\,:

\begin{equation}
ds^2 = A(r) dt^2 - B(r) dr^2 - r^2 d\Omega^2 \label{eq: schwartz}
\end{equation}
where $d\Omega^2 = d\theta^2 + \sin^2{\theta} d\varphi^2$ is the spherical 
line element (see 
\cite{arturo2,diego} for a discussion).  Combining the $00$\,-\,vacuum 
component and the trace of the field equations (\ref{eq:f-var2}) in 
absence of matter, we get the equation\,:

\begin{equation}
\label{master-low} f'(R) \left ( 3 \frac{R_{00}}{g_{00}} - R
\right ) + \frac{1}{2} f(R) - 3 \frac{f'(R)_{; 0 0}}{g_{00}} = 0 \
.
\end{equation}
Using 
Eq.(\ref{eq: frn}), Eq.(\ref{master-low}) reduces to\,:

\begin{equation}
\label{master-pla} R_{00}(r) = \frac{2 n - 1}{6 n} \ A(r) R(r) -
\frac{n - 1}{2 B(r)} \frac{dA(r)}{dr} \frac{d\ln{R(r)}}{dr} \ ,
\end{equation}
and the trace equation reads\,:

\begin{equation}
\Box{R^{n - 1}(r)} = \frac{2 - n}{3 n} R^n(r) \ . \label{eq:
tracebis}
\end{equation}
For $n = 1$, Eq.(\ref{eq: tracebis}) reduces to $R = 0$,
which, inserted into Eq.(\ref{master-pla}), gives $R_{00} = 0$ and
then the  standard Schwarzschild solution is recovered. Expressing 
$R_{00}$ and $R$ in terms of the metric (\ref{eq: schwartz}), 
Eqs.(\ref{master-pla}) and (\ref{eq: tracebis}) become a system of  
differential equations for  $A(r)$ and $B(r)$. A physically motivated 
hypothesis is assuming

\begin{equation}
A(r) = \frac{1}{B(r)} = 1 + \frac{2 \Phi(r)}{c^2} \label{eq:
avsphi}
\end{equation}
with $\Phi(r)$ the gravitational potential generated by a
pointlike mass $m$ at the distance $r$. A general solution is 
\cite{capo06,zakh06}:

\begin{equation}
\Phi(r) = - \frac{G m}{2 r} \left [ 1 + \left ( \frac{r}{r_c}
\right )^{\beta} \right ] \label{equ02}
\end{equation}
\noindent where $r_c$ is an arbitrary parameter, depending on the
typical scale of the considered system and $\beta$ is a universal
parameter:
\begingroup
\setlength{\abovedisplayskip}{0pt}
\setlength{\belowdisplayskip}{0pt}
\begin{equation}
\beta = \dfrac{12 n^2 - 7n - 1 - \sqrt{36 n^4 + 12 n^3 - 83 n^2 + 50n
+ 1} }{6 n^2 - 4n + 2}.
\label{equ03}
\end{equation}
\endgroup
Clearly the gravitational potential deviates from the 
Newtonian one because of the presence of the second term on the
right hand side. For $\beta = 0$, it is $n=1$ and the Newtonian
potential is fully recovered: the metric reduces to the 
Schwarzschild one and the weak field limit of GR is 
fully restored.

A comment is necessary at this point. The parameter $r_c$ is an 
integration constant depending on $\beta$. As shown in 
\cite{arturo}, it is related to the fact that any $f(R)$ power-law  
model possesses a Noether symmetry. In other words, it is a 
fundamental radius because strictly depends on the existence of the 
Noether symmetry. In this perspective, it is a further gravitational 
radius besides the Schwarzschild one and  depends on the mass and the size of the self gravitating system. Being this a central concept in this paper, details on the emergence of this further gravitational radius from Noether's symmetries are given in Appendix A.

The other important point  is related to the choice of the power-law action \eqref{eq: frn} that could appear non-natural in order to discuss deviations with respect to GR.  Being $n$ any real number, it is always possible to recast the  $f(R)$ power-law function as  
\begin{eqnarray}\label{LOGe}
f(R)\propto R^{1+\epsilon}\,.
 \end{eqnarray}
 If we assume small deviation with respect to GR, that is 
  $|\epsilon| \ll 1$,   it is possible to re-write
 a first-order Taylor expansion as 
 \begin{eqnarray}\label{LOG}
R^{1+\epsilon}&\simeq & R+\epsilon R {\rm log}R +O (\epsilon^2)\, .
 \end{eqnarray}
 In this way, one can 
control the magnitude of the corrections with respect to
the  Einstein gravity. 
Lagrangian  (\ref{LOG}) has been  investigated   from Solar System up to cosmological scales.  In particular, applications to  gravitational waves \cite{maurofelix},  binary star systems  \cite{mnrasfelix}, and neutron stars have been investigated \cite{farinelli}.
In the present paper, we consider the general case  $n\in \mathbb{R}$ because, as shown in Appendix A, the Noether symmetry exists for any $n$ and large deviations from GR  seem consistent with galactic dynamics as we will show below.

\begin{table}[ht!]
\caption{The data from the Baryonic Tully--Fisher relation of gas 
rich galaxies as a test of $\Lambda$CDM and MOND  
\cite{mcga11a}: $D$ - distance of the galaxy, $V_c$ - rotational 
velocity, $M*$ - mass of the stars, $Mg$ - mass of the gas.}
\begin{ruledtabular}
\begin{tabular}{lrcrrr}
Galaxy & $D$ & $V_c$ & log$M*$ & log$Mg$ \\
 & ($Mpc$) & ($km/s$) & ($Msun$) & ($Msun$) \\
\hline
DDO210 & 0.94 & 17 $\pm$ 4 & 5.88 $\pm$ 0.15 & 6.64 $\pm$ 0.20 \\
CamB & 3.34 & 20 $\pm$ 12 & 6.99 $\pm$ 0.15 & 7.33 $\pm$ 0.20 \\
UGC8215 & 4.5 & 20 $\pm$ 6 & 6.81 $\pm$ 0.15 & 7.45 $\pm$ 0.20 \\
DDO183 & 3.24 & 25 $\pm$ 3 & 7.24 $\pm$ 0.15 & 7.54 $\pm$ 0.20 \\
UGC8833 & 3.2 & 27 $\pm$ 4 & 6.94 $\pm$ 0.15 & 7.30 $\pm$ 0.20 \\
D564-8 & 6.5 & 29 $\pm$ 5 & 6.76 $\pm$ 0.20 & 7.32 $\pm$ 0.13 \\
DDO181 & 3.1 & 30 $\pm$ 6 & 7.26 $\pm$ 0.15 & 7.56 $\pm$ 0.20 \\
P51659 & 3.6 & 31 $\pm$ 4 & 6.67 $\pm$ 0.15 & 7.85 $\pm$ 0.20 \\
KK9824 & 7.83 & 35 $\pm$ 6 & 7.72 $\pm$ 0.15 & 7.93 $\pm$ 0.20 \\
UGCA92 & 3.01 & 37 $\pm$ 4 & 7.78 $\pm$ 0.15 & 8.32 $\pm$ 0.20 \\
D512-2 & 14.1 & 37 $\pm$ 7 & 7.58 $\pm$ 0.20 & 7.96 $\pm$ 0.06 \\
UGCA444 & 0.95 & 38 $\pm$ 5 & 7.34 $\pm$ 0.15 & 7.75 $\pm$ 0.20 \\
KK98251 & 5.6 & 38 $\pm$ 5 & 7.34 $\pm$ 0.15 & 8.02 $\pm$ 0.20 \\
UGC7242 & 5.4 & 40 $\pm$ 4 & 7.57 $\pm$ 0.15 & 7.78 $\pm$ 0.20 \\
UGC6145 & 7.4 & 41 $\pm$ 4 & 7.20 $\pm$ 0.15 & 7.56 $\pm$ 0.20 \\
NGC3741 & 3.0 & 44 $\pm$ 3 & 7.24 $\pm$ 0.15 & 8.45 $\pm$ 0.20 \\
D500-3 & 18.5 & 45 $\pm$ 6 & 6.97 $\pm$ 0.20 & 7.94 $\pm$ 0.05 \\
D631-7 & 5.5 & 53 $\pm$ 5 & 6.88 $\pm$ 0.20 & 8.29 $\pm$ 0.15 \\
DDO168 & 4.3 & 54 $\pm$ 3 & 8.07 $\pm$ 0.15 & 8.74 $\pm$ 0.20 \\
KKH11 & 3.0 & 56 $\pm$ 5 & 7.28 $\pm$ 0.15 & 7.85 $\pm$ 0.20 \\
UGC8550 & 5.1 & 58 $\pm$ 3 & 8.25 $\pm$ 0.37 & 8.46 $\pm$ 0.39 \\
D575-2 & 12.2 & 59 $\pm$ 7 & 7.63 $\pm$ 0.20 & 8.62 $\pm$ 0.07 \\
UGC4115 & 7.5 & 59 $\pm$ 6 & 7.77 $\pm$ 0.15 & 8.58 $\pm$ 0.20 \\
UGC3851 & 3.2 & 60 $\pm$ 5 & 8.45 $\pm$ 0.15 & 9.09 $\pm$ 0.20 \\
UGC9211 & 12.6 & 64 $\pm$ 5 & 8.12 $\pm$ 0.39 & 9.21 $\pm$ 0.41 \\
NGC3109 & 1.3 & 66 $\pm$ 3 & 7.41 $\pm$ 0.15 & 8.79 $\pm$ 0.20 \\
UGC8055 & 17.4 & 66 $\pm$ 7 & 8.09 $\pm$ 0.15 & 9.02 $\pm$ 0.20 \\
D500-2 & 17.9 & 68 $\pm$ 7 & 7.41 $\pm$ 0.20 & 9.06 $\pm$ 0.05 \\
IC2574 & 4.0 & 68 $\pm$ 5 & 8.94 $\pm$ 0.15 & 9.20 $\pm$ 0.20 \\
UGC6818 & 18.6 & 72 $\pm$ 6 & 9.22 $\pm$ 0.16 & 9.28 $\pm$ 0.20 \\
UGC4499 & 13.0 & 74 $\pm$ 3 & 8.75 $\pm$ 0.27 & 9.32 $\pm$ 0.29 \\
NGC1560 & 3.45 & 77 $\pm$ 3 & 8.70 $\pm$ 0.15 & 9.23 $\pm$ 0.20 \\
UGC8490 & 4.65 & 78 $\pm$ 3 & 8.36 $\pm$ 0.15 & 8.96 $\pm$ 0.20 \\
UGC5721 & 6.5 & 79 $\pm$ 3 & 8.17 $\pm$ 0.37 & 9.05 $\pm$ 0.39 \\
F565-V2 & 48. & 83 $\pm$ 8 & 8.30 $\pm$ 0.21 & 9.04 $\pm$ 0.24 \\
F571-V1 & 79. & 83 $\pm$  5 & 9.00 $\pm$ 0.19 & 9.33 $\pm$ 0.22 \\
IC2233 & 10.4 & 84 $\pm$ 5 & 8.96 $\pm$ 0.15 & 9.32 $\pm$ 0.20 \\
NGC2915 & 3.78 & 84 $\pm$ 10 & 7.99 $\pm$ 0.15 & 8.78 $\pm$ 0.20 \\
NGC5585 & 5.7 & 90 $\pm$ 3 & 8.98 $\pm$ 0.38 & 9.27 $\pm$ 0.40 \\
UGC3711 & 7.9 & 95 $\pm$ 3 & 8.92 $\pm$ 0.15 & 9.01 $\pm$ 0.17 \\
UGC6983 & 18.6 & 108 $\pm$ 3 & 9.53 $\pm$ 0.16 & 9.74 $\pm$ 0.20 \\
F563-V2 & 61. & 111 $\pm$ 5 & 9.41 $\pm$ 0.17 & 9.63 $\pm$ 0.21 \\
F568-1 & 85. & 118 $\pm$ 4 & 9.50 $\pm$ 0.18 & 9.87 $\pm$ 0.22 \\
F568-3 & 77. & 120 $\pm$ 6 & 9.62 $\pm$ 0.18 & 9.71 $\pm$ 0.22 \\
F568-V1 & 80. & 124 $\pm$ 5 & 9.38 $\pm$ 0.18 & 9.65 $\pm$ 0.22 \\
NGC2403 & 3.18 & 134 $\pm$ 3 & 9.61 $\pm$ 0.15 & 9.77 $\pm$ 0.20 \\
NGC3198 & 14.5 & 149 $\pm$ 3 & 10.12 $\pm$ 0.15 & 10.29 $\pm$ 0.20 \\
\end{tabular}
\end{ruledtabular}
\label{tab01}
\end{table}

\section{Observational constraints for $r_c$ from  BTF relation and circular velocity}
\label{sec_btf}

Starting from the above solution (\ref{equ02}), an excellent 
agreement between theoretical and observed rotation curves of low 
surface brightness galaxies has been obtained for $\beta$ = 0.817 
(see \cite{capo07}). A similar result has been achieved by Frigerio 
Martins and Salucci \cite{frig07} considering an enlarged sample 
of galaxies with respect to \cite{capo07}. Both these results can be 
framed into the BTF relation with the aim to show that the new 
fundamental gravitational radius $r_c$ can account for missing 
matter in galaxies.

Specifically, the empirical BTF relation is a universal relationship 
between the baryonic mass of a galaxy and its rotational velocity of 
the form $M_b\varpropto v_c^4$, which follows from the fact that 
luminosity $L$ traces baryonic mass $M_b$ through the mass-to-light 
ratio $\Upsilon$. The BTF relation can be recovered 
from $f(R)$ power-law gravity \cite{capo06,zakh06,zakh07,bork12}, 
but similar considerations can be developed for any $f(R)$ gravity 
model. As shown in \cite{capo07}, the BTF relation perfectly holds 
for low-surface brightness galaxies for $\beta$ = 0.817. 

Before considering the general discussion of BTF relation, we have to 
say that the Gauss theorem does not hold for $f(R)$ gravity because 
they predict the scaling of gravitational force other than $1/r^2$, 
as it is in the Newtonian case \cite{capo07}. But nevertheless, the 
Bianchi identities hold so the fundamental conservation law of any 
theory of gravity is guaranteed. Furthermore, according to the 
results below, it seems that this violation, in the case of 
BTF relation, has no dramatic consequence. Therefore, we will study 
the simplest point-like case where the modified gravitational 
potential $\Phi(r)$ generated by a point-like mass $M$ at the 
distance $r$ is given by Eq. \ref{equ02}.

Circular velocity of a point mass, in the $R^n$ gravity potential
(\ref{equ02}),  can be found in the standard way, that is
$v^2_c(r)=r\dfrac{d\Phi}{dr}$, which gives \cite{binney}:
\begingroup
\setlength{\abovedisplayskip}{0pt}
\setlength{\belowdisplayskip}{0pt}
\begin{equation}
v^2_c(r)=\frac{GM}{2r}\left[1+(1-\beta)\left(\frac{r}{r_c}
\right)^\beta\right] .
\label{equ04}
\end{equation}
\endgroup

For a detailed explanation of the relation for circular 
velocity $v_c$, see \cite{capo07}. There, it is described the low 
energy limit of power-law $f(R)$ models. We emphasize that the 
circular velocity definition that we are using here can be assumed as 
the proper one. This point needs further explanation.

As shown in \cite{salv12}, considering the Newtonian limit 
of $f(R)$  gravity and discarding higher order terms than 
$\mathcal{O}(2)$, the field equations for a perfect-fluid  
energy-momentum tensor  of dust ($p = 0$)  become:
\begin{eqnarray}
\label{field_eqs}
&&\nabla^2 \Phi-\frac{R^{(2)}}{2}-f''(0)\nabla^2 R^{(2)}\,= 
\,\mathcal{X}\rho\,,
\\
&&-3f''(0)\nabla^2 
R^{(2)}-R^{(2)}\,=\,\mathcal{X}\rho\,,\label{HOEQTRA}
\end{eqnarray}
where $\rho$ is the mass density,  $\mathcal{X}=\frac{8\pi G}{c^4}$ is the 
gravitational coupling, and $R^{(2)}$ is the Ricci scalar assumed up to the second order approximation. 
Due to the presence of $R^{(2)},$ these equations are of fourth-order.
For $f''(0)\,=\,0$, the standard second-order Poisson equation. 
$\nabla^2\Phi\,=\,4\pi G\rho$, is recovered and Eq.(\ref{HOEQTRA}) 
becomes the trace of standard GR in the weak field limit.

In general, the Poisson equation is different for $f(R)\neq 
R$, however, it is important to stress that the potential emerging 
as a solution, in our case that in Eq.(\ref{equ02}), must satisfy the 
geodesic equation that has the same form both in GR and in $f(R)$ 
gravity. Furthermore, the gravitational field is conservative (and 
irrotational) in both cases, then the equation $\boldsymbol{a} 
=-\nabla {\Phi}$, where ${\boldsymbol a}$ is the acceleration and ${\Phi}$ is 
the  potential, must hold for the standard Newtonian potential or 
other potentials derived in the Newtonian limit \cite{salv12}. This 
means that the circular velocity $v_c$, in the non-relativistic 
limit, can be obtained by $v^2_c(r)=r\dfrac{d\Phi}{dr}$. Eq. 
(\ref{equ04}) holds for  $f(R)\propto R^n$.
  
Finally, the derivation of Poisson equation, for general 
$f(R)$ gravity, is given by Zhao et al. \cite{zhao11}.

Let us now consider the fact that MOND can be recovered from $f(R)$ 
gravity in order to compare the 
predictions for BTF relation. In other words, it is possible to show 
that MOND is not only a phenomenological approach but it can be 
recovered as the weak field limit of Extended Theories of Gravity 
\cite{salv11}. Since this is a vital concept for the present paper, 
we will present a summary of the derivation reported in Ref. 
\cite{bern11}. We will proceed step by step  to demonstrate 
this basic concept for our approach.
\begin{itemize}
 \item[-] As reported in Appendix A,  the Noether symmetries select 
 a power-law  for $f(R)$ gravity. This is the only 
general form of $f(R)$ function presenting symmetries. In 
particular, we assume $f(\chi) = \chi^n$, after 
introducing the dimensionless quantity $\chi := L_M^2 R$, where $R$ 
is the Ricci scalar, $L_M$ is the length fixed by the parameters 
of the theory, and $n$ any real number.
 \item[-] The trace of  field Eqs. (\ref{eq:f-var2}) 
can be rewritten as 
\begingroup
\setlength{\abovedisplayskip}{0pt}
\setlength{\belowdisplayskip}{0pt}
\begin{equation}
f'(\chi) \, \chi  - 2 f(\chi) + 3 L_M^2  \, \Delta  f'(\chi) = 
 \frac{ 8 \pi G L_M^2 }{ c^4} T, 
\end{equation}
\endgroup
with the trace of the energy-momentum $T := 
T^\alpha_\alpha$. By substituting the power-law,  it becomes:
\begingroup
\setlength{\abovedisplayskip}{0pt}
\setlength{\belowdisplayskip}{0pt}
\begin{equation}
 \chi^n  \left(n - 2 \right) - 3 n L_M^2  \frac{ \chi^{(n-1)} }{ 
r^2 } \approx \frac{ 8 \pi G M L_M^2 }{ c^2 r^3}. \label{traccia}
\end{equation}
\endgroup
Here, we are assuming the weak field approximation with $d/d\chi\sim1/\chi$, $\Delta\sim -1/r^2$, and  matter density  
$\rho\sim M/r^3$.
The second term in the l.h.s. of Eq.(\ref{traccia}) is larger than the first if
\begin{equation}
Rr^2\leq \frac{3n}{2-n}\,.
\end{equation}
In this approximation, we can assume that the Ricci scalar 
corresponds to the Gaussian curvature and then $R\simeq R_c^{-2}$ 
where $R_c$ is the Gauss curvature radius. Immediately we have 
\begin{equation}
\label{regime}
 R_c >> r, 
 \end{equation}
 and then
\begingroup
\setlength{\abovedisplayskip}{0pt}
\setlength{\belowdisplayskip}{0pt}
\begin{equation}
R^{ (n-1)} \approx - \frac{ 8 \pi G M  }{ 3 n c^2 r L_M^{2 \left( n - 
1 \right) } }\,.
\end{equation}
\endgroup
 \item[-] At the second order, the Ricci scalar is
 \begin{equation}
 R=-\frac{2}{c^2}\nabla^2\Phi=\frac{2}{c^2}\nabla\cdot \boldsymbol{a}
 \end{equation}
that can be approximated as $R \approx -  2 \Phi/ (c^2 r^2)  \approx 
2 a/( c^2 r)$, with $\Phi$ the gravitational potential and $a$ the acceleration. This  gives:
\begin{eqnarray}
a &\approx& - \frac{ c^2 r }{ 2 L_M^2 }  \left( \frac{ 8 \pi G M  
}{ 3 n c^2 r  } \right)^{1/\left( n - 1 \right)}\nonumber\\ & 
\approx & - c^{\left( 2 n- 4  \right)/\left( n - 1 \right) } r^{ 
\left( n - 2 \right) / \left( n - 1 \right) } L_M^{-2} \left( G M  
\right)^{ 1 / \left( n - 1 \right) }, \nonumber
\end{eqnarray}
which converges to a MOND-like acceleration $a \propto
1 / r \) if \( n - 2 = - \left( n - 1  \right)$, that means
$n = 3 / 2$.
 \item[-] With this value of $n$, we get the  MOND relation 
\begingroup
\setlength{\abovedisplayskip}{0pt}
\setlength{\belowdisplayskip}{0pt}
\begin{equation}
a \approx - \frac{ \left( a_0 G M \right)^{1/2} }{ r }\,,
\end{equation}
\endgroup
reported in Ref. \cite{bern11}.
\end{itemize}
In other words, the weak field limit of $f(R)$ power-law gravity 
gives MOND as a particular case.

According to this derivation, the above characteristic length 
$r_c$ of $R^n$ gravity can be related to the MOND acceleration 
constant $a_0$ using the following expression \cite{bern11}:
\begingroup
\setlength{\abovedisplayskip}{0pt}
\setlength{\belowdisplayskip}{0pt}
\begin{equation}
r_c=\sqrt{\frac{GM}{a_0}} .
\label{equ05}
\end{equation}
\endgroup
Assuming that rotation curve (\ref{equ04}) is flat within the
measurement uncertainties at some finite radius $r_f$, i.e.
$v_c(r_f)\approx v_f$, then $r_f$ could be also related to a certain
MOND acceleration $a_f > a_0$ according to \cite{bern11}. This gives
\begingroup
\setlength{\abovedisplayskip}{0pt}
\setlength{\belowdisplayskip}{0pt}
\begin{equation}
r_f=\frac{\sqrt{a_0GM}}{a_f}=\frac{a_0}{a_f}\; r_c.
\label{equ06}
\end{equation}
\endgroup

Hence, from (\ref{equ04}), (\ref{equ05}), and (\ref{equ06}), the BTF 
relation of $R^n$ gravity expressed in terms of MOND accelerations 
is: 
\begingroup
\setlength{\abovedisplayskip}{0pt}
\setlength{\belowdisplayskip}{0pt}
\begin{equation}
M=\frac{4a_0 v^4_f}{Ga_f^2\left[1+(1-\beta)\left(\dfrac{a_0}{a_f}
\right)^\beta\right]^2} \, .
\label{equ07}
\end{equation}
\endgroup

\noindent We should point out that in the case of BTFR for spiral 
galaxies, McGaugh \cite{mcga11b} has shown that instead of standard 
MOND acceleration constant $a_0$ one should use a slightly 
different, empirically calibrated constant $a$ (where $a_0 = 0.8 
a$), while the formula is unchanged. Therefore, for our calculations 
we use the following expression:
\begingroup
\setlength{\abovedisplayskip}{0pt}
\setlength{\belowdisplayskip}{0pt}
\begin{equation}
M=\frac{4a v^4_f}{Ga_f^2\left[1+(1-\beta)\left(\dfrac{a}{a_f}
\right)^\beta\right]^2} \, .
\label{equ08}
\end{equation}
\endgroup

\begin{figure}[ht!]
\centering
\includegraphics[width=0.48\textwidth]{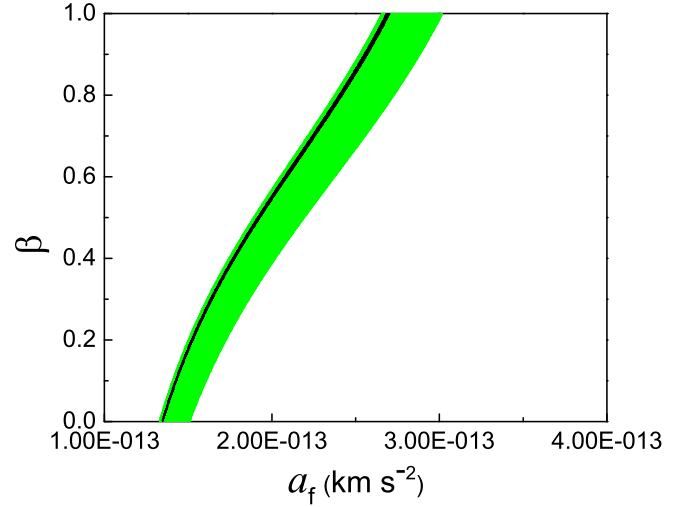}
\caption{Region in $(a_f,\;\beta)$ parameter space (shaded green 
area) where $41\leq A\leq 53\;
\mathrm{M}_{\odot}\,\mathrm{km}^{-4}\,\mathrm{s}^{4}$ according to
Eq. (\ref{equ08}). Black solid line represents the case when $A = 
51.4\; \mathrm{M}_{\odot}\,\mathrm{km}^{-4}\,\mathrm{s}^{4}$.}
\label{fig01}
\end{figure}

As it can be seen from eq. (\ref{equ08}), the slope of BTF relation 
in $R^n$ gravity is 4, which is the same as in the case of MOND:
 $M_b = Av_f^4$ where $A = 47\pm6 \;
\mathrm{M}_{\odot}\,\mathrm{km}^{-4}\,\mathrm{s}^{4}$.
Fig. \ref{fig01} shows the region in $(a_f,\;\beta)$ parameter space
where $41\leq A\leq 53\;
\mathrm{M}_{\odot}\,\mathrm{km}^{-4}\,\mathrm{s}^{4}$ according to
Eq. (\ref{equ08}). As it can be seen from Fig. \ref{fig01}, $R^n$
gravity universal constant $\beta$ is correlated with MOND
acceleration $a_f > a_0$.

\begin{figure*}[ht!]
\centering
\includegraphics[width=0.45\textwidth]{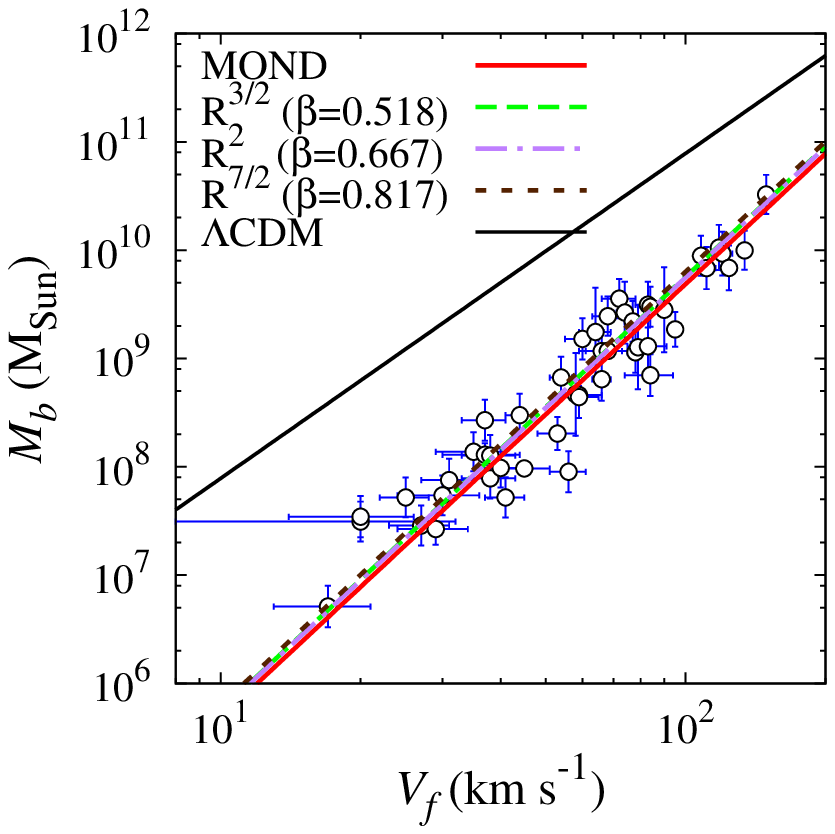}
\hspace{0.8cm}
\includegraphics[width=0.45\textwidth]{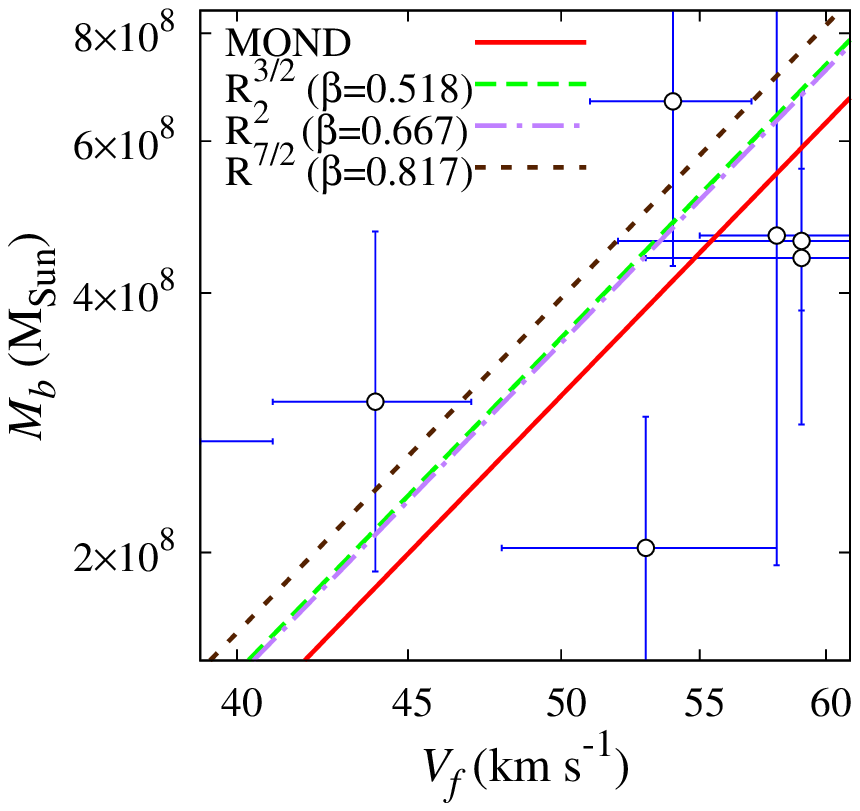}
\caption{\textit{Left:} Comparison between best fit BTF relations of 
gas-rich galaxies (for a sample of galaxies used in \cite{mcga11a}), 
in MOND, $R^n$ gravity for values of $n$ = 1.5, 2 and 3.5 
(corresponding $\beta$ are 0.518, 0.667 and 0.817, respectively) and 
$\Lambda$CDM. \textit{Right:} A zoomed part of the figure, for one 
small range of parameters. Notice: all values we calculated, except 
for open circles which are observed data from \cite{mcga11a}.}
\label{fig02}
\end{figure*}

In order to determine the best fit normalization, we adopted the
following values of universal constants in these two theories: 
$\beta=0.817$, obtained by fitting the Type Ia Supernovae Hubble 
diagram \cite{capo07} and $a_0 = 1.24 \times 10^{-13}$ km s$^{-2}$
\cite{mcga11a}, and then we fitted eq. 
(\ref{equ08}) to the observations of gas-rich galaxies given in 
Table 1 of \cite{mcga12}, considering MOND acceleration $a_f$ as a 
free parameter. The fit resulted with the following value of MOND
acceleration: $a_f(\beta = 0.817) = (2.56 \pm 0.06) \times 10^{-13}$ 
km s$^{-2}$, which yields the normalization of $51.4 \;
\mathrm{M}_{\odot}\,\mathrm{km}^{-4}\,\mathrm{s}^{4}$, being
within the error interval of the corresponding MOND estimate. 
Except value $n$ = 3.5 ($\beta$ = 0.817) for $R^n$ gravity we also 
investigated values of $n$ = 1.5 and 2 corresponding $\beta$ are 
0.518 and 0.667, respectively. In the same way like in the case 
of $\beta$ = 0.817, in order to determine the best fit normalization, we 
proceed with calculation. The fit resulted with the following value 
of MOND accelerations for $\beta$ = 0.518: $a_f(\beta = 0.518) = 
(2.02 \pm 0.05) \times 10^{-13}$ kms$^{-2}$, and for $\beta$ = 
0.667: $a_f(\beta = 0.667) = (2.30 \pm 0.06) \times 10^{-13}$ 
kms$^{-2}$, respectively.

A graphical comparison between the best fit BTF relations in $R^n$
gravity, MOND and $\Lambda$CDM is presented in Fig. \ref{fig02}. The 
data (open circles) are for a sample of galaxies from 
\cite{mcga11a}, and to make it more convenient for the readers, in 
Table \ref{tab01} we present data for gas rich galaxies that we used. 
MOND and $\Lambda$CDM lines are obtained using the relations 
given in \cite{mcga11a}. From Fig. \ref{fig02}, it can be seen that 
almost a perfect alignment is obtained between MOND and $R^n$ 
gravity predictions ($R^n$ gravity for values of $n$ = 1.5, 2 and 
3.5 corresponding $\beta$ are 0.518, 0.667 and 0.817, respectively). 
On the contrary, $\Lambda$CDM fit is in unsatisfactory agreement 
with observations. We can see that the range for $\beta$ from 0.5 to 
0.8 (corresponding to range of $n$ from 1.5 to 3.5) is in a good 
agreement with observations. For Fig. \ref{fig02}, we used data 
for a sample of galaxies which is used in \cite{mcga11a}, which we 
designate with open circles. We want to stress that these values 
give two constraints on parameter $\beta$ obtained from fitting 
observations using $R^n$ gravity, from the Fundamental Plane (FP) 
\cite{bork16} and from MOND \cite{bern11, mendoza1,  mendoza2, hidalgo, sobouti}.

By substituting the best fit value for $a_f$ in eq. (\ref{equ06}), 
one obtains:
\begingroup
\setlength{\abovedisplayskip}{0pt}
\setlength{\belowdisplayskip}{0pt}
\begin{equation}
r_f\approx\frac{r_c}{2}.
\label{equ09}
\end{equation}
\endgroup
The fact that the radii $r_f$ of all galaxies are constant when
expressed in units of the characteristic length $r_c$ demonstrates
the fundamental nature of $r_c$. Moreover, it is possible to make an
analogy between this situation and, for example, the radius of 
innermost stable circular orbit around a Schwarzschild black hole
which is always equal to 3 Schwarzschild radii, regardless of the
black hole mass. Therefore, $r_c$ represents a new fundamental 
gravitational radius which plays an analogues role in the case of 
weak gravitational field at galactic scales, as the Schwarzschild 
radius in the case of strong gravitational field in the 
vicinity of compact massive objects.

\section{Conclusions}
\label{sec_concl}

In this paper we used power-law $f(R)$ gravity to demonstrate the 
existence of a new fundamental gravitational radius. This radius 
plays an analog role, in the case of weak gravitational field at 
galactic scales, like the Schwarzschild radius in the case of strong 
gravitational field in the vicinity of compact massive objects. 
Such a feature emerges from Noether's symmetries that exist 
for any power-law $f(R)$ function. In particular, for \( f(R) 
\propto R^{3/2} \), the MOND acceleration regime is recovered. 
Using this new gravitational radius, $f(R)$ theories of gravity are 
able to explain the baryonic Tully--Fisher relation of gas-rich 
galaxies in a natural way and without need for dark matter 
hypothesis.

We can conclude that the range for $\beta$ from 0.5 to 0.8 
(corresponding to range of $n$ from 1.5 to 3.5) is in a good 
agreement with observations. This values represent two constraints on 
parameter $\beta$ obtained from fitting the FP and from MOND. In 
conclusion, we do not need any dark matter hypothesis in order to 
explain baryonic Tully-Fisher relation, and even more, $\Lambda$CDM 
fit is not in satisfactory agreement with observations.

We can conclude that $f(R)$ gravity can give a theoretical foundation 
both for  rotation curve of galaxies \cite{salucci1,salucci2} and even for the empirical  
BTF relation. We have to stress that obtained value for parameter 
$\beta$ from BTF differs from parameter $\beta$ at Solar System 
scales or from S2 star orbit scales \cite{bork12}. The reason for 
this result is that gravity is not a scale-invariant interaction and 
then it differs at galactic scales with respect to local scales. The 
constraints working at Solar System level for GR could not be simply 
extrapolated at any astrophysical and cosmological scale. However 
further and detailed investigations are necessary to confirm this 
statement and to reproduce the whole dark sector phenomenology at 
all scales.

\begin{acknowledgments}
S.C. acknowledges the support of INFN ({\it iniziative specifiche}
TEONGRAV and QGSKY). P.J., V.B.J. and D.B. wish to acknowledge the 
support by the Ministry of Education, Science and Technological 
Development of the Republic of Serbia through the project 176003. 
The authors also acknowledge the support of the Bilateral 
Cooperation between Serbia and Italy 451-03-01231/2015-09/1 
''Testing Extended Theories of Gravity at different astrophysical 
scales'', and of the COST Actions MP1304 (NewCompStar) and CA15117 
(CANTATA), supported by COST (European Cooperation in Science and 
Technology).
\end{acknowledgments}

\begin{appendix}
\section{The Noether Symmetry Approach}
\label{noether}

The fundamental feature on which are based the above results 
is the existence of a further gravitational radius related to a 
conserved quantity. In this Appendix, we briefly summarize the 
Noether Symmetry Approach that leads to the emergence of such a 
radius. 

As discussed in details in \cite{arturo,bern11}, the existence of a
Noether symmetry selects a power law form for the $f(R)$ function. 
  
Let us take into account 
an analytical $f(R)$ gravity. As discussed in 
Sec. \ref{sec_btf}, we can transform $f(R)\rightarrow f(\chi)$  
assuming $\chi : =L_M^2R$. With this transformation, we  deal with 
homogeneous dimensional quantities \cite{bern11}.

A point--like $f(\chi)$ Lagrangian is obtained by imposing the 
spherical symmetry into the action \eqref{equ01}. This means to 
reduce the infinite degrees of freedom of the field theory to a 
finite number. The approach consists in choosing suitable Lagrange 
multipliers into the action \cite{arturo,defelice}.

Let us assume a static spherically symmetric metric of the form 
\begin{equation}
  \mathrm{d}s^2 = A(r) c^2 \mathrm{d}t^2 - B(r) \mathrm{d}r^2 - C(r)
    \mathrm{d}\Omega^2,
\label{me2}
\end{equation}
where $ \mathrm{d}\Omega^2 := \mathrm{d}\theta^2 +
\sin^2{\theta} \mathrm{d}\varphi^2 $ is the angular component.  
A point--like  Lagrangian \( L \) for $f(\chi)$ is obtained by recasting
 action~\eqref{equ01} as
\begin{equation}
  A = - \frac{ c^3 }{ 16 \pi G L_M^2 } \int{ \left[ f(\chi) -
      \lambda (\chi - \bar{\chi}) \right] \sqrt{-g} \, \mathrm{d}^4x}\,.
\label{lma}
\end{equation}
Here $\lambda$ is a Lagrange multiplier related to $\bar{\chi} = 
L_M^2 \bar{R}$ expressed in terms of the metric~\eqref{me2}.
Due to the spherical symmetry, we can fix the functions  $B(r)$ or $C(r)$.  Both the assumptions are equivalent 
under a suitable change of coordinates \cite{arturo}. With this consideration in mind, the Ricci 
scalar can be expressed by two of the three functions in \eqref{lma}, i.e. 
\begin{equation}
  \bar{R} = \frac{A''}{A} + \frac{2 C''}{C} + \frac{A'C'}{AC} -
            \frac{{A'}^2}{2 A^2} - \frac{{C'}^2}{2 C^2} - \frac{2}{C}\,,
\label{ricci}
\end{equation}
where prime is the derivative with respect to $r$. Varying with
respect to \( \chi \) gives the Lagrange multiplier
\begin{equation}
  \lambda = \frac{ \mathrm{d} f(\chi) }{ \mathrm{d} \chi } := f_\chi\,,
\end{equation}
and then substituting into \eqref{lma} and discarding the boundary 
terms, the point--like Lagrangian reduces to
\begin{equation}
  \begin{split}
    \mathrm{L} = & - \frac{L_M^2}{\sqrt{A}} \left[ \frac{A
      f_\chi}{2C}{C'}^2 + f_\chi A' C' + C f_{\chi\chi}A' \chi' +
      \right. \\
    & 2 A f_{\chi\chi} C' \chi' \bigg] - \sqrt{A} \left[
      (2 L_M^2 + C \chi) f_\chi - C f \right]\,,
  \end{split}
\label{lag1}
\end{equation}
which is canonical in the configuration space $ \mathbb{Q}\equiv \{ 
A,C,\chi\} $ and tangent space velocities $ \mathbb{TQ}\equiv 
\{A',A, C', C, \chi', \chi\} $.
Assuming the regime given by  \eqref{regime} and the related weak 
field approximation, the last two terms of \eqref{lag1} are both 
much smaller than \( L_M^2 f_\chi \). This allows to rewrite 
Lagrangian~\eqref{lag1} as
\begin{equation}
  \begin{split}
    \mathrm{L} = & - \frac{L_M^2}{\sqrt{A}} 
      \left[ \frac{A f_\chi}{2C}{C'}^2 + f_\chi A' C' + 
      C f_{\chi\chi}A' \chi' + \right. \\
    & 2 A f_{\chi\chi} C' \chi' + 2 A \bigg]\,.
  \end{split}
\label{lag2}
\end{equation}
If Noether symmetries exist, they are  related to conserved quantities and cyclic variables 
that allow to reduce the dynamics
\cite{arnold,marmo,morandi}. Furthermore, the existence of a 
symmetry \textit{selects} the functional form of $f(\chi)$ gravity.

In general, a symmetry, and then a conserved quantity, exists 
if the Lie derivative of the Lagrangian \eqref{lag2} along a 
vector field \textbf{X} vanishes:
\begin{equation}
  \mathcal{L}_{\boldsymbol{\mathrm{X}}} \mathrm{L} = \alpha_i \nabla_{q_i}
  \mathrm{L} + \alpha'_i \nabla_{q'_i} \mathrm{L} = 0\,,
\label{lie}
\end{equation}
where $\, i=1,2,3,$ in the configuration space $\mathbb{Q}\equiv 
\{A,C,\chi\}$.
Solving the vector equation \eqref{lie} means to find out the functions
$\alpha_{i}$ which constitute the components of the Noether vector
\begin{equation}
\boldsymbol{\mathrm{X}}=\alpha_i\frac{\partial}{\partial q_i}+\alpha'_i\frac{\partial}{\partial q'_i}\,,
\end{equation}
defined on the tangent space $\mathbb{TQ}$.

Eq.\eqref{lie} gives also  the functional form of $f(\chi)$, see also 
\cite{prado,rubano,f(R)-cosmo,defelice}.

In the present case, a general form of Noether vector, related to the Killing equations of the model  \cite{townsend}, is 
\begin{gather}
  \alpha_1 = k_1 A + p_1, \nonumber \\
  \alpha_2 = k_2 C + p_2, \nonumber \\
  \alpha_3 = k_3 \chi + p_3,
\label{alpha}
\end{gather}
where $k_i$ and $p_i$ are constants. The  Lie 
condition \eqref{lie} is satisfied for 
\begin{equation}
  \boldsymbol{\alpha} = \biggl\{2(1-n)kA,\ 0,\ k\chi \biggr\},\quad f(\chi)=\chi^n,
\label{soluz}
\end{equation}
where $k$ is an integration constant and $n\in \mathbb{R}$. The 
related constant of motion $\Sigma_{0}$ is
\begin{equation}
  \begin{split}
    \Sigma_{0} &=  \alpha_i \nabla_{q_i'}{\mathrm L},  \\ 
&= L_M^2 n (n-1) k A^{-1/2} C \chi^{n-2} \left[2 (n-1) A \chi' 
    - A' \chi \right]. 
  \end{split}
\label{cm}
\end{equation}
In the case of GR, i.e. for \( n = 1 \), the Noether symmetry 
gives the conserved quantity \( \Sigma_0 = 2 r_\text{g} \), which is 
exactly the Schwarzschild radius. In the 
case of MOND, where equation~\eqref{cm} is valid for \( n = 3/2 \), 
\( C(r) = r^2 \) and, at the lowest order of perturbation, $A(r)=1 +
2\Phi/c^2 $, the constant of motion is given by
\begin{equation}
  \Sigma_0 = \frac{3}{2} k r_\text{g}^2 l_M\,.
\label{sigma-mond}
\end{equation}
Clearly this relation involves two characteristic lengths, 
a mass length-scale \( l_M \), that is the above $r_c$, and the Schwarzschild radius \( r_\text{g} \). In other words, the 
Noether symmetry gives a conserved quantity related to a further 
gravitational radius in the case $n \neq 1$.
\end{appendix}

\end{document}